\documentclass[10pt,conference]{IEEEtran}
\IEEEoverridecommandlockouts
\usepackage{cite}
\usepackage{amsmath,amssymb,amsfonts}
\usepackage{algorithmic}
\usepackage{graphicx}
\usepackage{textcomp}
\usepackage{xcolor}
\def\BibTeX{{\rm B\kern-.05em{\sc i\kern-.025em b}\kern-.08em
    T\kern-.1667em\lower.7ex\hbox{E}\kern-.125emX}}
    
\begin{document}

\title{A Review on Software Architectures for Heterogeneous Platforms}


\author{\IEEEauthorblockN{Hugo Andrade}
\IEEEauthorblockA{\textit{Department of Computer Science and Engineering} \\
\textit{Chalmers $|$ University of Gothenburg}\\
Gothenburg, Sweden \\
sica@chalmers.se}
\and
\IEEEauthorblockN{Ivica Crnkovic}
\IEEEauthorblockA{\textit{Department of Computer Science and Engineering} \\
\textit{Chalmers $|$ University of Gothenburg}\\
Gothenburg, Sweden \\
crnkovic@chalmers.se}
}

\maketitle

\begin{abstract}
The increasing demands for computing performance have been a reality regardless of the requirements for smaller and more energy efficient devices. Throughout the years, the strategy adopted by industry was to increase the robustness of a single processor by increasing its clock frequency and mounting more transistors so more calculations could be executed. However, it is known that the physical limits of such processors are being reached, and one way to fulfill such increasing computing demands has been to adopt a strategy based on heterogeneous computing, i.e., using a heterogeneous platform containing more than one type of processor. This way, different types of tasks can be executed by processors that are specialized in them. Heterogeneous computing, however, poses a number of challenges to software engineering, especially in the architecture and deployment phases. In this paper, we conduct an empirical study that aims at discovering the state-of-the-art in software architecture for heterogeneous computing, with focus on deployment. We conduct a systematic mapping study that retrieved 28 studies, which were critically assessed to obtain an overview of the research field. We identified gaps and trends that can be used by both researchers and practitioners as guides to further investigate the topic.
\end{abstract}


\section{Introduction}

The demands for computing performance keep increasing. Especially in the domain of cyber-physical systems, there is a large amount of data to be processed and critical requirements to be satisfied. Throughout the years, the hardware industry has aimed at increasing the processors' clock frequency in order to process more data. In this sense, the physical density of the chips has been increased with the addition of more and more transistors and therefore improving the capacity of the processing unit (PU). However, we are reaching the physical limit of processors with such strategy \cite{Brodtkorb2010}, uprising the need for a different way to continue increasing hardware performance in accordance to the also increasing software demands.

One way to handle such performance requirements is through heterogeneous computing \cite{Mittal2015}. It refers to the use of processors of different types in a computer system, such as CPUs, GPUs and FPGAs. In software engineering, the goal in employing heterogeneous computing is to decompose the software system into comprehensive kernels and assign data to processors that are specialized in them. For example, in a computer system containing one CPU and one GPU as an accelerator, the program control data may be processed by the CPU while multiple floating-point calculations are processed by the GPU. This strategy allows for better performance in high-demand systems, but at the same time requires a robust orchestration of hardware resources and the inherent software complexity.

A key aspect in heterogeneous computing is software deployment \cite{Andrade2016}, through which the mapping between software kernels and PUs is created. In addition to data types to be processed, there are several other attributes to be taken into consideration for the decomposition of the software system and allocation onto PUs. Critical aspects include, for instance, the proximity between PUs and bandwidth through which messages will be passed. Such a complex environment demands a software architecture that accounts for these diverse aspects while supporting software deployment on heterogeneous platforms. The software architecture must enable the software to take full advantage of the hardware resources, considering the different nature of the available processors. 

In this paper, we describe the conduction of a systematic mapping study that aims at investigating the state-of-the-art of software deployment on heterogeneous platforms, focusing on the architecture discipline. Our intention is to provide an overview of the research area, allowing both practitioners to acknowledge approaches and researchers to abide to opportunities for future research. This systematic mapping study gathers common practices while highlighting trends and gaps in research. 

This paper is derived from a larger study investigating further aspects of software deployment on heterogeneous platforms. Due to the importance of the architecture discipline within the context of this topic, we present our findings separately in the present paper.

The remainder of this paper is organized as follows. Section~\ref{sec:background} describes the background. Section~\ref{sec:researchmethod} presents the research method used in this study. Section~\ref{sec:results} provides the results after critically assessing the primary studies. In section~\ref{sec:discussion}, we discuss those results and reflect on their impact to the research area. Section~\ref{sec:threatstovalidity} describes the threats to the validity of this work. Section~\ref{sec:relatedwork} presents the related work. Finally, in section~\ref{sec:conclusion} we conclude presenting our final remarks.

\section{Background}
\label{sec:background}

The topic of heterogeneous computing has been increasingly studied over the past few years, mainly due to the general conclusions that the combination of different types of processors - rather than the choice between one over another - can bring performance improvements. In our experience conducting this systematic mapping study, we identified multiple definitions to common terms used in this area of research. Thus, it is important to clarify a few terms used in this paper in order to avoid misconceptions.

\textit{Heterogeneous platform}: Refers to a set of processing units of different types within a computing system. We found multiple studies that refer to this term in different ways. Besides meaning different processors, we found that this term also refers to platforms containing processors of the same type, but with different capacities. For instance, a system that includes 2 CPUs with a different number of cores and/or clock frequencies is often called heterogeneous. Another situation in which the term is commonly found is when the types and further characteristics of the processors are omitted, being discussed only the difference in capacity of the PUs. For example, strictly combinatorial problems that take into account a cost formula and a few performance attributes of the processors in order to determine the best deployment strategy. In this paper, we only consider systems that clearly and explicitly include processors of different types, such as CPUs, GPUs and FPGAs. 

\textit{Software deployment}: Refers to a stage within the software engineering process in which the (ready to be executed) software is placed onto the target hardware for execution. As we conducted this review, we realized that the activities performed in this stage are heavily influenced by activities in previous stages in the software process. For instance, we learned that one common way to realize deployment onto heterogeneous platforms is by using a development framework, which needs to be applied as soon as in the architecture phase. For this reason, we extend the concept of deployment to include all activities that are relevant throughout the software engineering process to successfully execute software onto a heterogeneous platform.

\textit{Software architecture}: Refers to a discipline within the software engineering process in which the structure of software is defined. It contains entities - typically components and connectors - that together represent the design of the system. Further, the software architecture also defines rules through which components communicate with each other. Although we sometimes mention the hardware architecture, as it is relevant to this topic of research, in this paper we focus primarily on software architecture.

Given the aforementioned concepts, the scope of our study sits primarily in the architecture design stage of the software engineering process. We do not set boundaries for investigation within a specific discipline, but are rather interested in the causes and effects that activities have towards the software architecture in heterogeneous computing environments.

\section{Research Method}
\label{sec:researchmethod}

As previously mentioned, this paper is part of a larger study that included aspects other than architecture in the investigation of software deployment on heterogeneous platforms. The process described here was conducted as part of such study, with the difference that, for this study, we selected only the papers referring to architecture and analyzed them separately. As the goal is to identify the state-of-the-art of software deployment on heterogeneous platforms, we performed a literature review in the form of a systematic Mapping Study (MS). MSs differ from classic Systematic Literature Reviews in their broadness and depth \cite{Budgen2008, KC07}. Rather than having a narrow focus on the investigation, in this study we aim at obtaining a broad overview of the research area through categorizing papers and aspects within them. 

This study followed the steps below, which were based on the guidelines proposed in \cite{Petersen2008}.

\begin{enumerate}
\item Definition of research question 
\item Conduction of search
\item Screening of papers
\item Keywording using abstracts
\item Data extraction and mapping process
\end{enumerate}

Prior to the definition of research questions, we composed a Review Protocol~\footnote{http://www.cse.chalmers.se/$\sim$sica/phd/mappingstudy} to thoroughly define the review process. The document serves as a guide during the review and includes information such as the motivation for a review, rationale to the research questions, inclusion/exclusion criteria and facets in which papers are categorized. All steps of the review were documented to allow traceability between them and enable reproducibility. Three researchers were involved in the processes of defining and conducting the review. Multiple meetings were held in order to align concepts, findings, and validate partial results that were obtained individually. 

In the following subsections, we describe the review steps.


\subsection{Research question}

From the goal of this study, we elaborated three research questions that cover aspects of interest within the topic of software deployment on heterogeneous platforms. The first research question refers to the \textit{\textbf{main concerns}} involved in software deployment on heterogeneous platforms. From this question we discovered the main reasons why software is deployed on heterogeneous platforms, as well as the issues that typically arise in the process. The second research question refers to the \textit{\textbf{approaches}} used to deploy software on heterogeneous platforms. This research question aimed at investigating the current state-of-the-art concerning activities, procedures, methods, approaches, and practices for deploying software on heterogeneous platforms. The third research question is the one we focus on this paper, and refers to the \textit{\textbf{architecture solutions}} for deploying software on heterogeneous platforms.

In other words, the main interest in this paper is related to practices within the architectural discipline that allow for software deployment on heterogeneous platforms. Thus, the following research question was formulated: 

\begin{itemize}
\item \textit{\textbf{Which architecture solutions enable/support deployment strategies for heterogeneous platforms?}}
\end{itemize}

With this research question we aim at exploring practices or standards that are used in the architecture level of a system containing a heterogeneous hardware platform. We considered any type of architectural solution that was reported to be used in such a heterogeneous context. We observed practices performed during the architecture design of a system, with focus on their implications to software deployment. In addition to answering the research question, we analyzed a number of aspects of each study in order to categorize them. Such categorization allows for the creation of a map of the research area, through which gaps and trends are visible.

\subsection{Conduction of search}

From the research questions, we extracted keywords and formulated the search string that served as input to the selected search engines. The search string is shown in Table~\ref{tab:searchstring} and was iteratively adjusted through a set of pilot studies until we obtained satisfactory results from the search engines. We defined the string based on the combination of key terms for the search: ``\textit{software}'' OR synonyms, with ``\textit{deployment}'' OR synonyms, with ``\textit{heterogeneous platforms}'' OR synonyms. When available, we used the ``advanced'' or ``expert'' search mode from the engine with an adapted version of the search string as input, in order to fulfill particular syntax requirements. We selected six digital libraries that include peer-reviewed studies and we judged to be the most relevant in the field of computer science and software engineering. 

\begin{table}[ht]
\footnotesize
\centering	
\caption{Search String}

	\begin{tabular}{|c|}
	    \hline
		      ``software'' \textbf{OR} ``program'' \textbf{OR} ``programs'' \textbf{OR} \\
		      ``application'' \textbf{OR} ``applications''  \\	    	 
	    \hline
	    	\textbf{AND} \\	    
	    \hline
	    	``deployment'' \textbf{OR} ``deploy'' \textbf{OR} ``deploying'' \textbf{OR} \\ 
	    	``installation'' \textbf{OR} ``install'' \textbf{OR} ``installing'' \textbf{OR} \\
		``allocation'' \textbf{OR} ``allocate'' \textbf{OR} ``allocating'' \\
	    \hline
	    	\textbf{AND} \\
	    \hline
	    	(``heterogeneous'' \textbf{OR} ``multiple'' \textbf{OR} ``hybrid'') \\ 
		\textbf{AND} \\
	    	(``platforms'' \textbf{OR} ``processing units'') 
	    	\\
	    \hline    
	\end{tabular} 

\label{tab:searchstring}
\end{table}


We searched the following search engines: ACM Digital Library\footnote{https://dl.acm.org/}, Engineering Village\footnote{https://www.engineeringvillage.com/}, IEEE Xplore\footnote{https://ieeexplore.ieee.org/}, ScienceDirect\footnote{https://www.sciencedirect.com/}, Scopus\footnote{https://www.scopus.com/} and Web of Science\footnote{https://www.webofknowledge.com/ }. The studies that were retrieved from the search engines were confronted with the pre-defined inclusion and exclusion criteria. These criteria were elaborated in order to reflect the objectives of the review and attest the relevance of the papers retrieved to this study.

\begin{itemize}
  \item \textit{Inclusion Criteria.} The papers must explore practice, theory, approaches or issues related to software deployment on heterogeneous platforms. We do not limit the types of processors that are discussed. When a study has been published in more than one venue, the most complete version was included. We consider full papers published in conferences, journals and workshops published up to (and including) 2017, written in English.

\item \textit{Exclusion Criteria.} Studies that do not address software deployment on heterogeneous platforms were excluded. Studies that mention software deployment, but do not discuss any type of method, activity, experience, or approach concerning means to deploy software were also excluded. We also excluded papers that refer to heterogeneous platforms in a sense other than a hardware containing more than one type of processor. This study does not cover heterogeneous distributed systems, e.g., high performance computers or Internet of the Things. We excluded studies that were only available as abstracts, PowerPoint presentations, tutorials, panels or demonstrations. Finally, short papers (three pages or less) were also excluded.
\end{itemize}

\subsection{Screening of papers}

\begin{figure*}[ht]
\centering
\resizebox{0.8\hsize}{!}{%
  \includegraphics{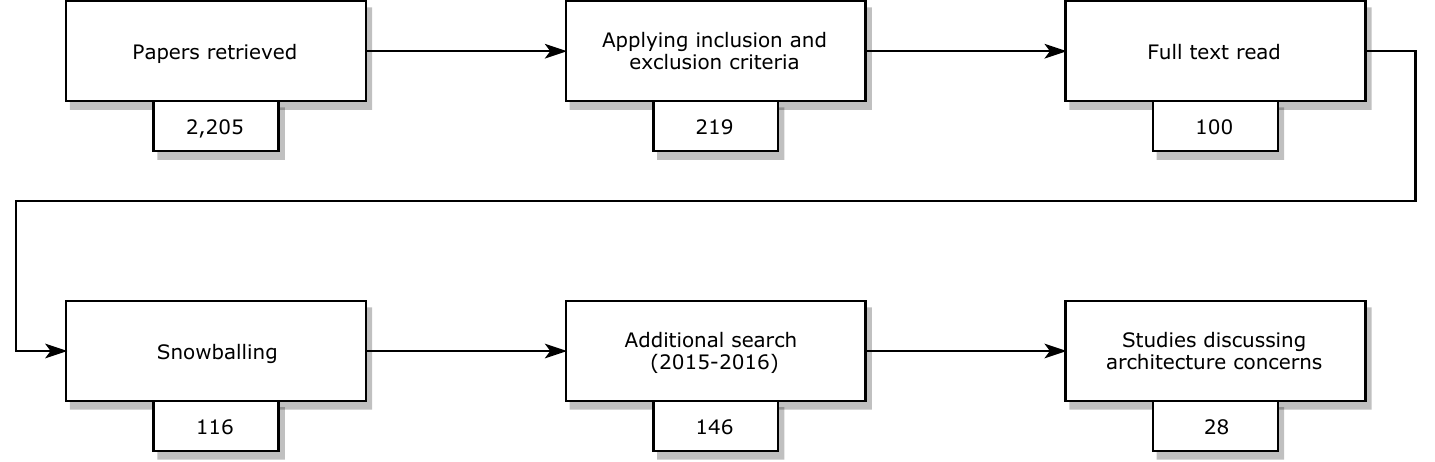}
}
\caption{Screening of papers}
\label{fig:screening}
\end{figure*}

The previously mentioned inclusion and exclusion criteria were considered to screen the retrieved papers and finally obtain the set of primary studies. The number of papers at each stage of the screening process is shown in Figure~\ref{fig:screening}. The first iteration considered studies that were published up to (and including) 2015. From the 2,205 results that were obtained from the search engines, 345 were excluded for being duplicates, PowerPoint presentations, PDFs with only a table of contents, documents referring to patents, publisher news, etc. The titles and abstracts of the remaining 1,860 were independently checked by two researchers, whose analysis resulted in 1,485 studies mutually marked for exclusion. The authors carried out rounds of discussion to solve disagreements regarding the inclusion or exclusion of the remaining 375 studies. Involving multiple researchers served as means to reduce bias and calibrate the screening process. These rounds resulted in 219 studies that were selected for full-text read and further evaluation of the inclusion/exclusion criteria.

On this stage, by only reading titles and abstracts, the actual meaning of the term ``heterogeneous'' was unclear for 79 studies. It was often necessary to check multiple sections of these papers to grasp what was meant when the term was used. Examples of meanings included but were not limited to: (i) a hardware platform with different types of processors; (ii) a hardware platform with two processors of the same type, but with different speeds; and (iii) a set of computers with different capacities. Since the scope of this MS only considers platforms containing processors of different types, it was important to check this parameter to determine which studies should have been excluded and which should have been included. From the 79 papers checked, 14 referred to the setup we were interested in. By reading full texts of the remaining 154 entries, we verified that 100 addressed deployment and provided answers to the RQs to some extent and fulfilled all inclusion criteria. In order to minimize the threat of missing relevant studies in the field, we also conducted the snowballing procedure \cite{Wohlin2014} and obtained an additional 16 studies, resulting in a total of 116 studies. A new search iteration was conducted in 2018, restricted to studies published from 2015 to 2017. Studies published in 2015 were also searched because we detected a few cases in which the search engines had not indexed them even after several months into 2016. This additional round followed a screening procedure that included a check for 2015 duplicates, the application of inclusion/exclusion criteria, full-text read, and snowballing, resulting in 31 studies. The results of the two iterations we combined to form the set of 146 primary studies in the main study. Finally, we identified 28 studies that discuss the topic of architecture and are subjects of evaluation in this work. They are hereby identified as P1, P2, ... , P28, and their titles, authors and years of publication are shown in Table~\ref{tab:primarystudies} (Appendix).

\subsection{Keywording using abstracts}

The titles, abstracts and keywords of the selected studies were submitted to a n-gram automated analysis, which results are presented later in this paper. We developed a script that processes the text and retrieves the most common 2-word and 3-word terms. These results were important for the authors to acknowledge common terminologies in the field. We also used the outcomes as hints concerning the directions which research has been taking in the field. 

\subsection{Data extraction and mapping process}

Once the papers were identified and common terms were observed, we proceeded to full-text reading. This phase involved two researchers independently reading, and then a third researcher who resolved conflicts in understanding and categorizing studies. For each entry, in addition to the information that addressed the RQs, we collected data that allowed the studies to be classified into a scheme. The classification scheme takes into account both directly extracted data (e.g., number of citations) and information that depends on the reader's interpretation (e.g., research type classification). We present the classification and the outcomes of this MS in the following section. 

The goal of the data extraction process was to collect relevant data from the selected studies. Such data includes evidence that (i) allowed the classification of studies into the pre-set facets (e.g., contribution type), and (ii) contributed to some extent in providing answers to the RQs. As the studies were being analyzed, we searched for parts of the text that would address the research questions and sub-questions, updating the spreadsheets accordingly. A new category was created whenever the reasoning behind a particular text fragment did not match the already existing categories. 

\section{Results}
\label{sec:results}

This section presents the results of the study, starting with an overview of the research area through a classification scheme. We show the distribution of papers according to their publication years and type of research that was conducted. Then, we discuss the types of processors found in the reported studies and describe the results of the n-gram analysis on titles, abstracts and keywords. Finally, we list the main purpose of the included papers.

\subsection{Classification scheme}

\begin{figure}[ht]
\centering
\resizebox{1\hsize}{!}{%
  \includegraphics{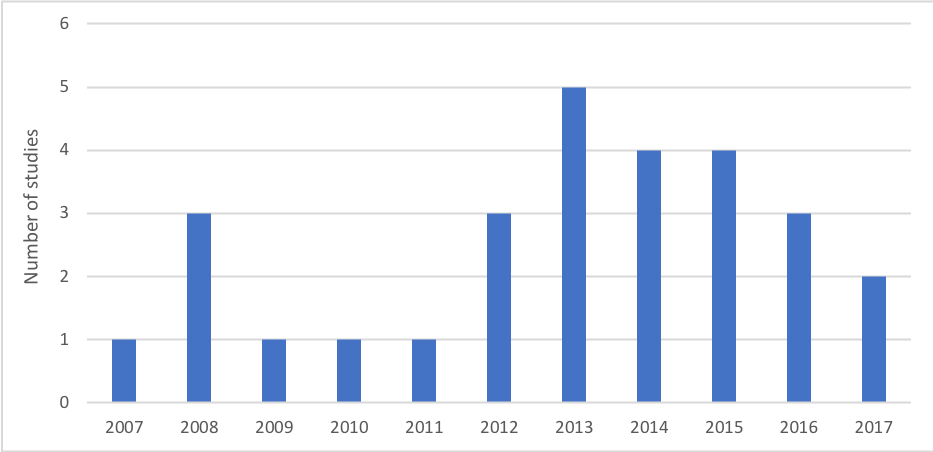}
}
\caption{Number of publications throughout the years}
\label{fig:years}
\end{figure}

\textit{Publication years.} The search for papers was not restricted to either a pre- or a post-defined publication time frame. As shown in Figure~\ref{fig:years}, the included papers were published within 2007-2017, which indicates that the research activity in the field is reasonably recent. We also observe a slight increase in the number of publications from the year 2012 when compared to the five previous years. The growth is probably motivated by the increasing interest on the topic of heterogeneous computing, triggered by the high demands for performance on several domains.

\textit{Number of citations.} In Table~\ref{tab:citations}, we present the 10 most cited primary studies, as of October 2018. The numbers were collected at the Scopus since all primary studies were indexed in this digital library. Out of these studies, 3 were published in Journals, 2 in Conferences, and 5 in Symposiums or Workshops. The most cited paper, P12, describes a heterogeneous platform containing FPGAs, GPUs and CPUs using a MapReduce framework.

\begin{table*}[t]
  \caption{Studies with the most citations according to Scopus (as of Oct, 2018)}
  \label{tab:citations}
  \begin{center}
  \resizebox{\textwidth}{!}{%
    \begin{tabular}{c|l|c|l} 
      \textbf{Citations} & \textbf{Study ID and Title} & \textbf{Year} & \textbf{Format} \\
      \hline
      83 & [P12] Axel: A Heterogeneous Cluster with FPGAs and GPUs & 2010 & Symposium  \\
      45 & [P16] Coordinating the use of GPU and CPU for improving performance of compute intensive applications & 2009 & Conference  \\
      35 & [P1] A Compiler and Runtime for Heterogeneous Computing & 2009 & Symposium  \\
      24 & [P24] dOpenCL: Towards uniform programming of distributed heterogeneous multi-/many-core systems & 2013 & Journal  \\
      15 & [P17] Design and initial performance of a high-level unstructured mesh framework on heterogeneous parallel systems & 2013 & Journal  \\
      13 & [P23] FPGA-GPU-CPU Heterogeneous Architecture for Real-time Cardiac Physiological Optical Mapping & 2012 & Conference  \\
      11 & [P11] Automatic synthesis of embedded SW for evaluating physical implementation alternatives from UML/MARTE models supporting memory space separation & 2014 & Journal  \\
      8 & [P19] Dynamic Reconfiguration of Tasks Applied to an UAV System Using Aspect Orientation & 2008 & Symposium  \\
      6 & [P3] A Federated Simulation Environment for Hybrid Systems & 2007 & Workshop  \\
      6 & [P5] A Scheduling and Runtime Framework for a Cluster of Heterogeneous Machines with Multiple Accelerators & 2015 & Symposium  \\
    \end{tabular}}
  \end{center}
\end{table*}

\textit{Research type classification.} In order to observe the type of research that has been conducted in the field, we followed the classification proposed in \cite{Wieringa2005}. As shown in Table~\ref{table:facetstypes}, the wide majority of the included studies conducted research aiming at proposing solutions to a given problem. Only one paper referred to evaluation research, aiming at assessing an existing technique. These numbers indicate that the level of maturity in the area is not high, being the priority the proposal of solutions to problems, rather than reporting experiences, validating approaches or evaluating existing solutions. 

\begin{table*}
\begin{scriptsize}
\caption{Research Type Facet as proposed in \cite{Wieringa2005}} 
\label{table:facetstypes} 
\begin{center}
\resizebox{15cm}{!}{%
	\begin{tabular}{|p{1.3cm}|p{9cm}|p{1.5cm}|}
	 	\hline
	 		\bfseries Types & \bfseries Description & \bfseries Number of papers
	 		 \\ \hline \hline
	 			Solution Proposal
				& A solution for a problem is proposed, the solution can be either novel or
				a significant extension of an existing technique. The potential benefits
				and the applicability of the solution is shown by a small example or a good
				line of argumentation. 
                & 27 \\                
			\hline
				Evaluation Research
				& Techniques are implemented in practice and an evaluation of the technique
				is conducted. Implementation of the technique is shown in practice
				(solution implementation) and the consequences of the implementation in
				terms of benefits and drawbacks (implementation evaluation) are
				demonstrated. 
                & 1 \\
			\hline
				Experience Papers
				& Experience papers explain what and how something has been done in
				practice. It has to be the personal experience of the author.
                & 0 \\	
			\hline
				Philosophical Papers
				& These papers sketch a new way of looking at existing things by
				structuring the field in form of a taxonomy or conceptual framework.
                & 0 \\
			\hline
				Opinion \  \  \ Papers
				& These papers express the personal opinion of somebody whether a certain
				technique is good or bad, or how things should be done. They do not rely
				on related work and research methodologies.
                & 0 \\
			\hline
				Validation Research
	 			& Techniques investigated are novel and have not yet been implemented in
	 			practice. Techniques used are, for example, experiments i.e. work done in
	 			the lab. 
                & 0 \\ 		
		\hline
	\end{tabular}}
	\end{center}
	\end{scriptsize}
\end{table*}

\textit{Types of processors.} We identified that 15 out of the 28 included studies (53.5\%), discussed heterogeneous hardware platforms consisting of CPU in combination with GPU. In 4 other cases, the authors considered platforms containing CPU, GPU and FPGA. In 3 studies, the authors reported platforms consisting of CPU + FPGA. In 3 studies, the platform consisted of CPU + GPU + other type of processor, such as DSP or MIC. Other less common combinations were found in 3 studies, such as the combination of general purpose GPU, FPGA and MIC processors as reported in \cite{Riebler2016}. The dominance of the CPU + GPU has also been identified through our n-gram analysis, as described next.

\textit{N-gram analysis on titles, abstracts and keywords.}  The titles, abstracts and keywords of the included studies were gathered in a text file and automatically analyzed in order to discover commonly used terms. The script performed both 2-gram and 3-gram analysis on the text, disregarding common natural language stop-words and irrelevant results such as the words "keyword" or "conference". The results in Table~\ref{tab:ngram} show that CPUs and GPUs are common terms used together, followed by terms related to quality attributes such as energy efficiency and performance. We can also observe a few domain-specific terms and challenges related to the allocation of tasks onto heterogeneous platforms.

\begin{table*}[t]
\begin{scriptsize}
  \caption{N-gram analysis on titles, abstracts and keywords}
  \label{tab:ngram}
  \begin{center}
  \resizebox{13cm}{!}{%
    \begin{tabular}{l c | l c}
      \multicolumn{2}{c|}{\textbf{2-gram}} & \multicolumn{2}{c}{\textbf{3-gram}} \\
      \hline
      \multicolumn{1}{c}{\textit{terms}} & \multicolumn{1}{c|}{\textit{count}} & \multicolumn{1}{c}{\textit{terms}} & \multicolumn{1}{c}{\textit{count}} \\
      \hline
      ``heterogeneous'', ``platforms'' & 7 & ``CPU'', ``+'', ``GPU'' & 4 \\
      ``CPUs'', ``GPUs'' & 7 & ``flexibility'', ``explore'', ``computational'' & 3 \\
      ``energy'', ``efficiency'' & 7 & ``multicore'', ``CPUs'', ``GPUs'' & 3 \\
      ``heterogeneous'', ``systems'' & 7 & ``application'', ``timing'', ``constraints'' & 3 \\
      ``embedded'', ``systems'' & 6 & ``parallel'', ``executable'', ``patterns'' & 3 \\
      ``runtime'', ``system'' & 5 & ``unmanned'', ``aerial'', ``vehicles'' & 3 \\
      ``system'', ``performance'' & 5 & ``task'', ``allocation'', ``decisions'' & 3 \\
      ``energy'', ``consumption'' & 5 & ``timing'', ``constraints'', ``design'' & 3 \\
      ``optical'', ``mapping'' & 5 & ``race'', ``condition'', ``detection'' & 3 \\
      ``unmanned'', ``aerial'' & 4 & ``role'', ``task'', ``allocation'' & 3 \\
    \end{tabular}}
  \end{center}
  \end{scriptsize}
\end{table*}

\textit{Main purpose of included papers.} To capture the main purpose of the studies, we focused on phrases that are typically included in the abstract, introduction and conclusion defining the main purpose of a given study. The identified purposes were diverse. Nine out of the 28 studies had the main purpose to propose a framework, algorithm, implementation or tool. On the other hand, 7 papers aimed at proposing a solution related to the problem of load balancing, including workload adjustment and resource management approaches. We identified 5 studies that aimed at either discussing or proposing solutions directly referring to architectural concerns: P4 \cite{Andrade2012}, P11 \cite{Posadas2014}, P22 \cite{Lai2017}, P23 \cite{Meng2012} and P28 \cite{Riebler2016}. 

In P4, the authors propose a framework that allows application domain experts to design the system. It includes specification of non-functional requirements and a hardware-software co-design environment that allows for a dynamic mapping using different operational scenarios.

In P11, the authors propose a methodology that enables the association between functional components in a given UML model to specific memory spaces. In this sense, through their approach it is possible to automatically allocate functional codes to different resources. From a UML/MARTE standard model, the approach allows for an exploration of different allocation possibilities for software components.

In P22, the authors propose an emulation tool that considers hardware information such as cache, memory and inter-processor communication attributes. Through hardware profiling, the approach provides a centralized interface for adding new accelerators in the emulation tool and detecting race conditions and performance analysis.

In P23, the authors propose a real-time architecture for systems in the health domain (cardiac optical mapping). It includes an optical mapping partitional analysis, and an experimentation setup featuring an NVIDIA GPU and a Xilinx FPGA.

In P28, the authors propose an architecture that allows for an automatic identification of \textit{hotspots} in the application code, at runtime, and generates corresponding binary code to target the specific accelerator. The solution uses a just-in-time compiler that works in collaboration with a resource management mechanism for dispatching applications onto the heterogeneous platform.

\subsection{Which architecture solutions enable/support deployment strategies for heterogeneous platforms?}

In the following subsections, we present the answer to the main research question through two points of view: architectural styles and architectural principles.

\subsubsection{Architectural styles} 
Refer to principles that define a family of such systems in terms of a pattern of substructural organization \cite{Garlan1993}. In other words, the term is often associated with patterns that respect a set of rules to facilitate and standardize the software system's structure and communication.

\paragraph{Layered architectures}

One solution to handle heterogeneity at the architectural level is using the layer pattern. In P1, for instance, the message passing is orchestrated by a dedicated communication layer that allows different processors (senders/receivers) to be aware of the other parties' desired data format. When the communication channels are implemented using such layered architecture strategy, developers may be able to avoid low-level memory copy and managing memory explicitly. However, the authors report that these changes come at the cost of decreasing OS and virtual machine portability. 

Another study that uses a layered approach is P16, which implementing an event executor layer that isolates the user provided code from the specific hardware concerns. The mapping between threads and hardware devices occurs at runtime by consulting a dedicated device scheduler. 

Further, in P27, the authors propose a component architecture consisting of five layers: component, ccaffeine framework, deployment, resource management and heterogeneous platforms. The resources management layer basically models and monitors the resources, providing resource status information to the deployment layer. In turn, the deployment layer creates a deployment strategy to satisfy pre-defined requirements and hardware characteristics.

\paragraph{Pipelined architectures}

In P3, the authors propose a performance-oriented environment that focuses on applications that are represented as general data flow graphs. The applications are expressed in a specific language as dataflow graphs. The approach bases the allocation strategy on the simulation of executing these graphs. The authors highlight the importance of simulation by using examples of application deployment on FPGAs.

Another pipeline-oriented approach is presented in P23, in which separate entities encapsulate computation groups. The application domain to which the architecture is proposed demands continuous execution of the system with a camera input.

\paragraph{Master-slave architectures}
In P13, the authors propose and evaluate an architecture based on the master-slave principle. It supports multiple allocation policies and workload adjustment techniques that are able to cope with load balancing problems. The approach basically establishes a relationship in which the slave(s) provide the master with relevant information for allocation of tasks, such as their processing speed.

\subsubsection{Architectural principles} 
Refer to practices within the software engineering process that aid in the design of software architectures. These principles define the baseline structure and constraints the architectural design, as discussed next. 

\paragraph{Separation of concerns}
The architecture design approach presented in P4 makes use of a design space exploration technique and is inspired by the Y-chart approach. The Y-chart approach proposes a deliberate separation of concerns related to the following aspects: application specification, platform model, and mapping between them.

In P5, the architecture design separates computation units from communication units using the concept of bulk synchronous computing. Their approach takes a task-graph previously defined by the application to the set of resources. Then, the load is balanced, the data exchange is abstracted and reduced, and the latency is hidden by overlapping computation and communication aspects.

\paragraph{Standardized architectures}
A number of studies discuss the use of a dedicated architecture solution for heterogeneous systems. In P7, for instance, the authors follow the guidelines and standards of hardware and software proposed by the HSA foundation\footnote{http://www.hsafoundation.com/}. One of the most prominent decisions in such architecture is the elimination of CPU-GPU data transfer overhead by designing principles that allow these processors to share the same data.

\paragraph{Aspect-oriented architectures}
In P19 and in P20, the architectures are defined following aspect-oriented principles. The task allocation strategies are defined based on the profiling of each task in different hardware scenarios. In this sense, several elements of the application are affected by the results of the task-resource mapping definition.

\paragraph{Dedicated communication structures}
Since communication is a critical aspect in heterogeneous systems, one common architectural design solution is to include a dedicated entity to handle communication between different processing units. In P14, the communication buses are annotated with non-functional properties, which later considered in the allocation process. In P21, the authors propose an architectural solution based on a middleware that enables communication via a dedicated proxy. It uses queues between programs written in different languages and amongst the heterogeneous processors.

Regarding hardware structures, most studies reported PCIe bus for communication, i.e., P2, P10, P12, P17, and P24.

\section{Discussion}
\label{sec:discussion}

The original search retrieved a very large number of papers, and thus represented a rather difficult process to identify relevant studies. We advocate the use of such a generic search string, because when the term "architecture" appeared in the search string on our pilot studies, papers were omitted since architectural concerns may be implicit. In terms of volume of research, we believe the number of papers on the topic is rather low when compared to the 146 originally retrieved in the broader topic. 

Along the conduction of search, we encountered a large number of papers discussing hardware concerns. Those papers were not included since the focus of our study was to focus on the software issues, and more specifically architectural design concerns. Another interesting finding is that approaches are heavily based on existing frameworks, such as OpenCL, which are arguably not easy to use. This represents a need for further approaches, methods and techniques that don't necessarily rely on standardized solutions and their inherent limitations. 

It is still very difficult to deploy software on some hardware platforms, such as FPGAs. The lack of software infrastructure and architectural solutions limits the popularity of heterogeneous systems using such types of processors. 

From the research community, there is a possibility that software architecture may not be the main concern, as in multiple cases practitioners are attempting to realize heterogeneous platforms according to requirements. In this sense, there are opportunities to put effort into new solutions that can be derived from existing architectural principles.

A few patterns were identified; but in general, architectural patterns might be realized in a high level of abstraction, while the papers identified in this review are dealing with low level problems. The development of systems aiming for execution at heterogeneous platforms can be improved if architectures on a high level of abstraction can be taken into account. Further, the styles highlighted previously give emphasis to communication (master-slave, pipeline), which is a fundamentally important aspect on the type of systems discussed in this paper. Finally, the most prominent architectural style that has been identified through this study is the architecture based on layers. The systems reported on the primary studies apply such strategy in order to abstract the heterogeneity caused by the underlying hardware.

Another interesting discussion is about the absence of service oriented architectures (SOA). The styles identified are mostly technology-oriented, instead of covering more loosely distributed principles. The mapping study focused on specific computation units, and SOA is used on distributed systems, where heterogeneity is typically defined on software level, instead of on hardware or computational level. It would be an interesting question to address in the future: how are the styles described in a higher abstraction level (SOA), and therefore make a connection between concerns and the heterogeneous executable units.

The focus of this work was on heterogeneous platforms, unlike heterogeneous systems, such as high performance computing and Internet of the Things. This paper did not address these types of systems. It might be that for such systems different types of architectural design solutions exist.

\section{Threats to Validity}
\label{sec:threatstovalidity}

The threats to validity of this work are mostly related to the search and data extraction processes. In systematic reviews and mapping studies, there is a possibility that researchers fail to retrieve all relevant papers in a given field. It can be that some published papers that discuss the investigated topic are neglected in the screening process due to search engines limitations, or human error. We reduced the possibility of missing such papers by conducting a process that is strictly systematic, reproducible and includes well-defined criteria for selecting studies. The entire review process was extensively discussed, validated and executed by two or even three researchers in order to reduce individual bias. Further, the search was conducted in multiple points in time, in order to cover papers that were possibly not yet available on databases, i.e., when the search is conducted in January, there is a high possibility that papers accepted in the end of the previous year were still not indexed.

Regarding the extraction process, we attempted to read the papers thoroughly in the search for information that addressed the topic of our investigation. Due to the large amounts of text, it is possible that relevant information was neglected. In order to mitigate such risk, we collected several attributes of the papers, in terms of meta-data, including their main purpose. This allowed us to categorize the papers and more easily discuss and validate among the researchers involved. We believe that by covering the main purpose of each paper, the core research idea and intention are captured, and therefore we obtain a reasonable overview of the research field.

\section{Related Work}
\label{sec:relatedwork}

In \cite{Mittal2015}, the authors thoroughly investigated heterogeneous computing techniques through a survey, including both software and hardware aspects. Their work includes approaches for workload partitioning and their uses against system performance and energy consumption requirements. The study reports an in-dept categorization of techniques that are used throughout the development of heterogeneous computing systems, such as programming languages, development frameworks and tools. However, their survey is limited to CPU-GPU environments. As shown in the findings of our study, CPU-GPU platforms represent today the majority of heterogeneous computing platforms. There is a variety of approaches that can be used when developing systems to be deployed on such platforms. On the other hand, we believe that other types of processors, such as FPGAs and DSPs are also gaining importance in industry and will soon become more common solutions in heterogeneous computing. FPGAs, for instance, are capable of high computing power despite the present difficulties in developing software to be executed on them. In the future, we believe that more tools and approaches will be available to decrease the upfront cost of implementing systems for this type of processors. 

Further, in \cite{Brodtkorb2010}, the authors conducted a study that aimed at describing and analyzing the state-of-the-art in heterogeneous computing. They investigated hardware, software tools and algorithms used to develop systems that include processors of different types, such as CPUs, GPUs and FPGAs. The authors extensively describe the concerns related to developing systems for heterogeneous platforms, and included programming languages for CPUs, GPUs and FPGAs. However, the term \textit{architecture} often referred to the hardware characteristics of each processor type, and their impact on developing systems. Our work differs from theirs in the sense that we focus on software architectures and their implications to deployment on heterogeneous platforms. We restricted our scope to the software engineering process and how the software architecture design supports the deployment on platforms that are heterogeneous.

\section{Conclusion}
\label{sec:conclusion}

The potential of heterogeneous computing is starting to be recognized by the community as one solution to achieve better performance. This approach poses a number of challenges especially on the software side, which is required to handle the complexity of multiple types of architectures that will process data. One key aspect of such environment is the software architecture, which orchestrates processing and communication by defining rules and enabling requirements to be satisfied. 

In this paper, we conducted a systematic mapping study on software architectures for heterogeneous computing. We searched for literature to discover the state-of-the-art approaches in the field. The search was followed by a critical analysis of the studies and the identification of gaps and trends that can be explored in the future.

We found that a number of architectural design principles are being used in order to implement such heterogeneous systems. However, we identified that only 5 out of the 28 studies had their main purpose to propose methods specifically for software architecture design in heterogeneous systems. This represents the low maturity level of the field and highlights the need for further investigations.

As future work, we intend to further investigate how the complexity is dealt with on the architectural level and propose software tools to be incorporated in the software engineering process that will increase the feasibility of heterogeneous computing.

\section*{Acknowledgment}

We would like to thank our colleague Jan Schr\"{o}der, who contributed to this work by helping in the screening of papers and participating in discussions.

\bibliography{IEEEabrv,library}

\begin{thebibliography}{10}
\providecommand{\url}[1]{#1}
\csname url@samestyle\endcsname
\providecommand{\newblock}{\relax}
\providecommand{\bibinfo}[2]{#2}
\providecommand{\BIBentrySTDinterwordspacing}{\spaceskip=0pt\relax}
\providecommand{\BIBentryALTinterwordstretchfactor}{4}
\providecommand{\BIBentryALTinterwordspacing}{\spaceskip=\fontdimen2\font plus
\BIBentryALTinterwordstretchfactor\fontdimen3\font minus
  \fontdimen4\font\relax}
\providecommand{\BIBforeignlanguage}[2]{{%
\expandafter\ifx\csname l@#1\endcsname\relax
\typeout{** WARNING: IEEEtran.bst: No hyphenation pattern has been}%
\typeout{** loaded for the language `#1'. Using the pattern for}%
\typeout{** the default language instead.}%
\else
\language=\csname l@#1\endcsname
\fi
#2}}
\providecommand{\BIBdecl}{\relax}
\BIBdecl

\bibitem{Brodtkorb2010}
\BIBentryALTinterwordspacing
A.~R. Brodtkorb, C.~Dyken, T.~R. Hagen, J.~M. Hjelmervik, and O.~O. Storaasli,
  ``State-of-the-art in heterogeneous computing,'' \emph{Sci. Program.},
  vol.~18, no.~1, pp. 1--33, Jan. 2010. [Online]. Available:
  \url{http://dx.doi.org/10.1155/2010/540159}
\BIBentrySTDinterwordspacing

\bibitem{Mittal2015}
\BIBentryALTinterwordspacing
S.~Mittal and J.~S. Vetter, ``A survey of cpu-gpu heterogeneous computing
  techniques,'' \emph{ACM Comput. Surv.}, vol.~47, no.~4, pp. 69:1--69:35, Jul.
  2015. [Online]. Available: \url{http://doi.acm.org/10.1145/2788396}
\BIBentrySTDinterwordspacing

\bibitem{Andrade2016}
H.~Andrade, ``Investigating software deployment on heterogeneous platforms,''
  in \emph{2016 13th Working IEEE/IFIP Conference on Software Architecture
  (WICSA)}, April 2016, pp. 272--276.

\bibitem{Budgen2008}
D.~Budgen, M.~Turner, P.~Brereton, and B.~Kitchenham, ``Using mapping studies
  in software engineering,'' in \emph{Proceedings of PPIG}, vol.~8, 2008, pp.
  195--204.

\bibitem{KC07}
\BIBentryALTinterwordspacing
B.~Kitchenham and S.~Charters, ``{Guidelines for performing Systematic
  Literature Reviews in Software Engineering},'' Keele University and Durham
  University Joint Report, Durham, UK, Tech. Rep., Jul. 2007. [Online].
  Available: \url{http://community.dur.ac.uk/ebse/biblio.php?id=51}
\BIBentrySTDinterwordspacing

\bibitem{Petersen2008}
\BIBentryALTinterwordspacing
K.~Petersen, R.~Feldt, S.~Mujtaba, and M.~Mattsson, ``Systematic mapping
  studies in software engineering,'' in \emph{Proceedings of the 12th
  International Conference on Evaluation and Assessment in Software
  Engineering}, ser. EASE'08.\hskip 1em plus 0.5em minus 0.4em\relax Swinton,
  UK, UK: British Computer Society, 2008, pp. 68--77. [Online]. Available:
  \url{http://dl.acm.org/citation.cfm?id=2227115.2227123}
\BIBentrySTDinterwordspacing

\bibitem{Wohlin2014}
\BIBentryALTinterwordspacing
C.~Wohlin, ``Guidelines for snowballing in systematic literature studies and a
  replication in software engineering,'' in \emph{Proceedings of the 18th
  International Conference on Evaluation and Assessment in Software
  Engineering}, ser. EASE '14.\hskip 1em plus 0.5em minus 0.4em\relax New York,
  NY, USA: ACM, 2014, pp. 38:1--38:10. [Online]. Available:
  \url{http://doi.acm.org/10.1145/2601248.2601268}
\BIBentrySTDinterwordspacing

\bibitem{Wieringa2005}
\BIBentryALTinterwordspacing
R.~Wieringa, N.~Maiden, N.~Mead, and C.~Rolland, ``Requirements engineering
  paper classification and evaluation criteria: A proposal and a discussion,''
  \emph{Requir. Eng.}, vol.~11, no.~1, pp. 102--107, Dec. 2005. [Online].
  Available: \url{http://dx.doi.org/10.1007/s00766-005-0021-6}
\BIBentrySTDinterwordspacing

\bibitem{Riebler2016}
H.~Riebler, G.~Vaz, C.~Plessl, E.~M.~G. Trainiti, G.~C. Durelli, E.~D. Sozzo,
  M.~D. Santambrogio, and C.~Bolchini, ``Using just-in-time code generation for
  transparent resource management in heterogeneous systems,'' in \emph{2016
  IEEE 2nd International Forum on Research and Technologies for Society and
  Industry Leveraging a better tomorrow (RTSI)}, Sept 2016, pp. 1--5.

\bibitem{Andrade2012}
H.~A. Andrade, A.~Ghosal, K.~Ravindran, and B.~L. Evans, ``A methodology for
  the design and deployment of reliable systems on heterogeneous platforms,''
  in \emph{2012 International Conference on Reconfigurable Computing and
  FPGAs}, Dec 2012, pp. 1--7.

\bibitem{Posadas2014}
\BIBentryALTinterwordspacing
H.~Posadas, P.~Peñil, A.~Nicolás, and E.~Villar, ``Automatic synthesis of
  embedded \{SW\} for evaluating physical implementation alternatives from
  uml/marte models supporting memory space separation,'' \emph{Microelectronics
  Journal}, vol.~45, no.~10, pp. 1281 -- 1291, 2014, dCIS’12 Special Issue.
  [Online]. Available:
  \url{http://www.sciencedirect.com/science/article/pii/S0026269213002607}
\BIBentrySTDinterwordspacing

\bibitem{Lai2017}
\BIBentryALTinterwordspacing
C.-K. Lai, C.-W. Yeh, C.-H. Tu, and S.-H. Hung, ``Fast profiling framework and
  race detection for heterogeneous system,'' \emph{J. Syst. Archit.}, vol.~81,
  no.~C, pp. 83--91, Nov. 2017. [Online]. Available:
  \url{https://doi.org/10.1016/j.sysarc.2017.10.010}
\BIBentrySTDinterwordspacing

\bibitem{Meng2012}
P.~Meng, M.~Jacobsen, and R.~Kastner, ``Fpga-gpu-cpu heterogenous architecture
  for real-time cardiac physiological optical mapping,'' in \emph{2012
  International Conference on Field-Programmable Technology}, Dec 2012, pp.
  37--42.

\bibitem{Garlan1993}
D.~Garlan and M.~Shaw, ``An introduction to software architecture,'' in
  \emph{Advances in Software Engineering and Knowledge Engineering}.\hskip 1em
  plus 0.5em minus 0.4em\relax Publishing Company, 1993, pp. 1--39.

\end{thebibliography}

\newpage

\appendix 

Table~\ref{tab:primarystudies}: List of primary studies included in this paper.

(next page)

\begin{center}
\begin{scriptsize}
\begin{table*}[t]
  \centering
  \begin{tabular}{p{17cm}cr}
  \hline
   \textbf{Study ID, Title, and Authors} & \textbf{Year}\\
   \hline
    \textbf{[P1]} \textbf{A Compiler and Runtime for Heterogeneous Computing} & 2012 \\ Auerbach, Joshua and Bacon, David F. and Burcea, Ioana and Cheng, Perry and Fink, Stephen J. and Rabbah, Rodric and Shukla, Sunil \\ \hline
    \textbf{[P2]} \textbf{A domain-specific language to facilitate software defined radio parallel executable patterns deployment on heterogeneous architectures} & 2014 \\ L. J. Mohapi and S. Winberg and M. Inggs \\ \hline
     \textbf{[P3]} \textbf{A Federated Simulation Environment for Hybrid Systems} & 2007 \\ Gayen, Saurabh and Tyson, Eric J. and Franklin, Mark A. and Chamberlain, Roger D. \\ \hline
     \textbf{[P4]} \textbf{A methodology for the design and deployment of reliable systems on heterogeneous platforms} & 2012 \\ H. A. Andrade and A. Ghosal and K. Ravindran and B. L. Evans \\ \hline
     \textbf{[P5]} \textbf{A Scheduling and Runtime Framework for a Cluster of Heterogeneous Machines with Multiple Accelerators} & 2015 \\ T. Beri and S. Bansal and S. Kumar \\ \hline
     \textbf{[P6]} \textbf{A user mode CPU–GPU scheduling framework for hybrid workloads} & 2016 \\ Wang, Bin and Ma, Ruhui and Qi, Zhengwei and Yao, Jianguo and Guan, Haibing \\ \hline
     \textbf{[P7]} \textbf{Accelerating DynEarthSol3D on tightly coupled CPU-GPU heterogeneous processors} & 2015 \\ Ta, Tuan and Choo, Kyoshin and Tan, Eh and Jang, Byunghyun and Choi, Eunseo \\ \hline
     \textbf{[P8]} \textbf{An extended model for multi-criteria software component allocation on a heterogeneous embedded platform} & 2013 \\ Ivan Svogor and Ivica Crnkovic \\ \hline
     \textbf{[P9]} \textbf{An FPGA-based architecture for embedded systems performance acceleration applied to Optimum-Path Forest classifier} & 2017 \\ Wendell F.S. Diniz, Vincent Fremont, Isabelle Fantoni, Eurípedes G.O. \\ \hline
     \textbf{[P10]} \textbf{Architecture Aware Resource Allocation for Structured Grid Applications: Flood Modelling Case} & 2015 \\ V. Saxena and T. George and Y. Sabharwal and L. V. Real \\ \hline
     \textbf{[P11]} \textbf{Automatic synthesis of embedded SW for evaluating physical implementation alternatives from UML/MARTE models supporting memory space separation} & 2014 \\ Héctor Posadas and Pablo Peñil and Alejandro Nicolás and Eugenio Villar \\ \hline
     \textbf{[P12]} \textbf{Axel: A Heterogeneous Cluster with FPGAs and GPUs} & 2010 \\ Tsoi, Kuen Hung and Luk, Wayne \\ \hline
     \textbf{[P13]} \textbf{Biological sequence comparison on hybrid platforms with dynamic workload adjustment} & 2013 \\ F. M. Mendonca and A. C. M. A. d. Melo \\ \hline
     \textbf{[P14]} \textbf{Component allocation optimization for heterogeneous CPU-GPU embedded systems} & 2014 \\ G. Campeanu and J. Carlson and S. Sentilles \\ \hline
     \textbf{[P15]} \textbf{Consolidating Applications for Energy Efficiency in Heterogeneous Computing Systems} & 2013 \\ J. Zhang and H. Wang and H. Lin and W. C. Feng \\ \hline
     \textbf{[P16]} \textbf{Coordinating the use of GPU and CPU for improving performance of compute intensive applications} & 2009 \\ G. Teodoro and R. Sachetto and O. Sertel and M. N. Gurcan and W. Meira and U. Catalyurek and R. Ferreira \\ \hline
     \textbf{[P17]} \textbf{Design and initial performance of a high-level unstructured mesh framework on heterogeneous parallel systems} & 2013 \\ Mudalige, G.R. and Giles, M.B. and Thiyagalingam, J. and Reguly, I.Z. and Bertolli, C. and Kelly, P.H.J. and Trefethen, A.E. \\ \hline
     \textbf{[P18]} \textbf{dOpenCL: Towards uniform programming of distributed heterogeneous multi-/many-core systems} & 2013 \\ P. Kegel and M. Steuwer and S. Gorlatch \\ \hline
     \textbf{[P19]} \textbf{Dynamic Reconfiguration of Tasks Applied to an UAV System Using Aspect Orientation} & 2008 \\ E. d. Freitas and A. P. D. Binotto and C. E. Pereira and A. Stork and T. Larsson \\ \hline
     \textbf{[P20]} \textbf{Dynamic Self-Rescheduling of Tasks over a Heterogeneous Platform} & 2008 \\ A. P. D. Binotto and E. P. Freitas and M. Götz and C. E. Pereira and A. Stork and T. Larsson \\ \hline
     \textbf{[P21]} \textbf{Enhanced Energy Efficiency with the Actor Model on Heterogeneous Architectures} & 2016 \\ Hayduk, Y. and Sobe, A. and Felber, P. \\ \hline
     \textbf{[P22]} \textbf{Fast profiling framework and race detection for heterogeneous system} & 2017 \\ Cheng-Kung Lai, Chih-Wei Yeh, Chia-Heng Tu, Shih-Hao Hung \\ \hline
     \textbf{[P23]} \textbf{FPGA-GPU-CPU Heterogenous Architecture for Real-time Cardiac Physiological Optical Mapping} & 2012 \\ P. Meng and M. Jacobsen and R. Kastner \\ \hline
     \textbf{[P24]} \textbf{Real-time task reconfiguration support applied to an UAV-based surveillance system} & 2008 \\ A. P. D. Binotto and E. P. de Freitas and C. E. Pereira and A. Stork and T. Larsson \\ \hline
     \textbf{[P25]} \textbf{Resource-awareness on heterogeneous MPSoCs for image processing} & 2015 \\ Paul, Johny and Stechele, Walter and Oechslein, Benjamin and Erhardt, Christoph and Schedel, Jens and Lohmann, Daniel and Schr\"{o}der-Preikschat, Wolfgang and Kr\"{o}hnert, Manfred and Asfour, Tamim and Sousa, \'{E}ricles and Lari, Vahid and Hannig, Frank and Teich, J\"{u}rgen and Grudnitsky, Artjom and Bauer, Lars and Henkel, J\"{o}rg \\ \hline
     \textbf{[P26]} \textbf{Runtime Resource Management in Heterogeneous System Architectures: The SAVE Approach} & 2014 \\ G. C. Durelli and M. Pogliani and A. Miele and C. Plessl and H. Riebler and M. D. Santambrogio and G. Vaz and C. Bolchini \\ \hline
     \textbf{[P27]} \textbf{Scheduling multi-paradigm and multi-grain parallel components on heterogeneous platforms} & 2011 \\ Y. Peng and C. Zhao and S. Yao and S. Li and Y. Chen \\ \hline
    \textbf{[P28]} \textbf{Using just-in-time code generation for transparent resource management in heterogeneous systems} & 2016 \\ Riebler, H. and Vaz, G. and Plessl, C. and Trainiti, E. M. G. and Durelli, G. C. and Del Sozzo, E. and Santambrogio, M. D. and Bolchini, C. \\ \hline
  \end{tabular}
  \caption{List of primary studies discussing architecture concerns}
  \label{tab:primarystudies}
\end{table*}
\end{scriptsize}
\end{center}

\end{document}